\documentclass[prl,twocolumn,showpacs]{revtex4}
\usepackage{epsfig}
\begin{document}
%
%
%
\title{
Strong Correlations Between Fluctuations and Response in Aging Transport
}
\author{E. Barkai\\
Department of Physics,
Bar Ilan University, Ramat-Gan 52900 Israel
}
\begin{abstract}
{ 
Once the problem of ensemble averaging is removed, correlations
between the  response
of a single molecule 
to an external driving field $F$, 
with the history of fluctuations of the particle, become detectable.
Exact analytical theory for the continuous time random walk and
numerical simulations for the quenched trap model give
the behaviors of the correlation
between fluctuations of the displacement
in the aging period $(0,t_a)$, 
and the response to bias 
switched on at time $t_a$. 
In particular in  the dynamical phase where the models exhibit
aging we find finite correlations
even in the asymptotic
limit $t_a \to \infty$, while in the non-aging phase the correlations
are zero in the same limit. Linear response theory gives
a simple relation between these correlations and the fractional
diffusion coefficient. 
}
\end{abstract}

\maketitle

 Originally the fluctuation--dissipation theorem was
formulated by Einstein using simple Brownian motion, i.e. the
well known Einstein relation between the diffusivity of an ensemble
of identical  particles
and their mobility \cite{Kubo}.  
Modern optical  techniques enable experimentalists
to track {\em  single molecules} 
undergoing either normal diffusion, or non Markovian
diffusion,  and in some cases anomalous diffusion
in a very wide variety of Physical and biological conditions 
\cite{BarkaiRev,Zumofen,Weitz,Princeton}. 
An issue over looked so far are the correlations
between the response of individual particles to an external
bias and their history of fluctuations.
 We will soon define these correlations
precisely, however for the time being consider
chemically identical single molecules under going a diffusion process 
in some random
medium. In the time period $(0,t_a)$, called the aging period,
different particles may exhibit
different fluctuations, 
for example if the particles sample different realizations
of disorder in the system.
Then an external driving field is switched on at time $t_a$. We realize
that the response of different particles might be correlated with 
with their diffusivity in the aging period. For example if a particle
happens to sample a region in space with deep traps in the aging period,
it is expected to diffusive slowly and then also
respond weakly compared with another
particle which happened to sample relatively shallow traps. 
However if the aging period is long 
$t_a \to \infty$ and the process ergodic, we may expect
that diffusivity of the particles in the aging period
becomes de-correlated 
from their response, since the particles had enough time to sample
the full random energy landscape in which they are moving.  

 Here we investigate the correlations between diffusion (fluctuations)
and mobility (response), found when single particle trajectories are analyzed.
We use two well known models:-
the Montroll-Weiss  continuous time random
walk (CTRW) model \cite{BouchaudREV,Metzler} 
and Bouchaud's quenched trap model \cite{WEB}.  These models
are known to exhibit anomalous diffusion \cite{BouchaudREV,Metzler}
and aging 
\cite{WEB,Monthus,Rinn,ACTRWPRL,Grigolini,Sokolov1,Sokolov2},
i.e. the
ensemble averaged response to an external field switched on
at time $t_a$ depends on the aging time 
(see details below). We  show that in the phase
where the models exhibit aging  
and {\em even in the
limit where $t_a \to \infty$} 
correlations between fluctuations and response
are finite, in complete contrast to non-aging dynamics. 
Besides the theoretical interest in correlations in aging transport,
the existence of such correlations can be used in principle to predict
which members of an ensemble of diffusing particles will
respond strongly (or weakly) to a driving force. Our work also
indicates that data analysis of the aging fluctuation-dissipation
relations must be made with care, 
since the response and the fluctuations are generally correlated. 

There are many examples of systems where such correlations might become
important, we mention the recent experiments on the diffusion 
of single LacI repressor 
proteins on DNA \cite{Princeton}, where a wide distribution
of diffusion coefficients of the proteins was found. This distribution
is likely due to the random DNA sequence the single protein explores.
Since the single molecules have widely distributed   
diffusion constants, their response to an external field
is likely correlated with their history of diffusion, e.g. particles
with small (or large)
diffusion  constants
have a weak (strong) response respectively. 
Other single particle experiments
of  micro-beads diffusing in actin networks \cite{Weitz}
exhibit power law waiting
times and anomalous diffusion, very much reminiscent of
the trap and CTRW models. Hence these single particle
systems are 
candidates for the investigation of the new correlations 
we consider in this manuscript, correlations which go beyond
the Einstein relations between the ensemble
average diffusion coefficient and mobility.  


{\em Model 1}  We consider the well known one dimensional
CTRW on a lattice 
\cite{BouchaudREV,Metzler,ACTRW,Flom}.
The lattice spacing is $a$ and the jumps are to nearest
neighbors only.
Waiting times between jump events are independent
identically distributed random variables with a common
probability density function (PDF) 
$\psi(\tau)$.
After waiting the particle  has a probability
$1/2 + h/2$ or $1/2-h/2$ to jump to the right or
left respectively. 
In the aging period $(0,t_a)$ $h=0$ and the
particles follow an  unbiased motion, while in
the response period $t_a<t$ the bias is  $0<h<1$.  
The total measurement time is $t=t_a + t_r$ where 
$t_r$ is called the response time.
To define the response one has to define the field  which
is responsible for the bias $h$. For example, if the particle
is coupled to a thermal heat bath with temperature $T$,
and  driven by a uniform force field $F$, 
standard detailed balance conditions give
$h=a F/ 2  k_b T$, 
when $h<<1$  
\cite{ACTRW,Bertin}. 
We consider later the generic case 
\begin{equation}
\psi(\tau) \sim {A \tau^{ -(1 + \alpha) } \over |\Gamma\left(- \alpha\right)|}
\label{eqpower}
\end{equation}
when $\tau \to \infty$ and $0<\alpha<1$, $A>0$. Specific values of
$\alpha$ for a wide range of physical systems and models are given in
\cite{BouchaudREV,Metzler}. For example for the
annealed version of the trap model $\alpha=T/T_g<1$ \cite{Monthus}. 
In this case the average waiting time
is infinite. 

The position of the particle at time $t_a + t_r$ is 
$ X= X_a + X_r$,
where $X_a$ $(X_r)$ is the displacement in the aging (response) 
periods respectively. 
More specifically
$X_a = \sum_{i = 1} ^{n_a} \Delta x_i ^{(a)}, \ \  
X_r = \sum_{i=1} ^{n_r} \Delta x_i ^{(r)},$
where $\Delta x_i ^{(a)}$
and $\Delta x_i ^{(r)}$
are the random jump lengths (of length $a$)
in the aging $(0,t_a)$ and response periods $(t_a,t_a+t_r)$ respectively.
While $n_a$ and $n_r$  are the random number of jumps in the aging  and
response
periods respectively.

We investigate the 
correlation function
$\langle \left( X_a \right)^2 X_r \rangle$
which is a measure for the correlation
between the fluctuations in the aging period $(X_a)^2$,
and the response to the driving force switched on at time $t_a$,
$X_r$. 
We define a  dimensionless fluctuation-response (${\rm FR}$) parameter 
\begin{equation} 
{\rm FR}(t_a,t_r) = { \langle \left( X_a \right)^2 X_r \rangle\over  \langle \left( X_a \right)^2\rangle \langle X_r \rangle} - 1,
\label{Eq04}
\end{equation}
which is equal zero when correlations vanish.
For the CTRW under investigation one can show that
$\langle \left( X_a \right)^2 X_r \rangle=h a^3 \langle n_a n_r \rangle$
and
\begin{equation}
{\rm FR}(t_a,t_r) ={\langle n_a n_r  \rangle\over \langle n_a \rangle \langle n_r \rangle} - 1.
\label{Eq07}
\end{equation}
Thus $\langle n_a n_r \rangle$,
 the correlations of the number of steps in the aging period
with the number of steps in the  response period 
gives 
a measure for the correlations
between the fluctuations in the displacement  in the aging period and 
the response to the bias.

Let  
$P_{t_a,t_r}(n_a,n_r)$ be the probability of making $n_a$ jumps
in the aging period and $n_r$ jumps in the response period. 
Knowledge of this function is needed for the calculation
of the 
${\rm FR}$ parameter and other high order correlation functions \cite{Mukamel}
which we will discussed in a future publication.
 The paths with $n_a$ $(n_r)$ jump events in the aging period (response period)
 clearly
satisfy $t_{n_a} < t_a < t_{n_a + 1}$ and 
$(t_{n_a + n_r} <t_r + t_a < t_{n_a + n_r +1})$ respectively,
where the subscript $n$ in $t_n$ is for the jump number.  
Hence 
$$ P_{t_a,t_r} \left( n_a,n_r\right) = $$
\begin{equation}
\langle I \left( t_{n_a} < t_a < t_{n_a + 1}\right) I \left( t_{n_a + n_r} < t_r + t_a < t_{n_a + n_r + 1}\right) \rangle
\label{Eq11}
\end{equation}
where $I(x)=1$ if the event in the parenthesis is true otherwise
it is zero. Us-usual we make use of  the double Laplace transform
 $t_a \to u$  and $t_r \to s$ 
of $P_{t_a,t_r}\left( n_a ,n_r\right)$, 
$ P_{u,s}\left( n_a , n_r \right)=
\int_0 ^\infty {\rm d} t_r e^{ - s t_r}  \int_0 ^\infty {\rm d} t_a e^{ - u t_a}  P_{t_a,t_r}\left(n_a,n_r\right). $
For the sake of space we do not discuss the details of the calculation
(which will be published later).
We use the model assumption of independent identically distributed
waiting times, namely the renewal property of the CTRW, and
find
\begin{equation}
 P_{u,s}\left( n_a ,n_r=0\right) = { \hat{\psi}^{n_a} \left( u \right) \over s} \left[ { 1 - \hat{\psi}(u) \over u} - { \hat{\psi}(s) - \hat{\psi}(u) \over u-s} \right],
\label{Eq17}
\end{equation}  
while for $n_r \ge 1$ 
\begin{equation}
P_{u,s} \left( n_a , n_r \right) = { \hat{\psi}^{n_a} \left( u \right) \hat{\psi}^{n_r - 1} \left( s \right) \over s \left( u - s \right) } \left[ 1 - \hat{\psi}\left( s \right) \right] \left[ \hat{\psi} \left( s \right) - \hat{\psi} \left( u \right) \right].
\label{Eq18}
\end{equation}
In Eqs. (\ref{Eq17},\ref{Eq18}) 
$\hat{\psi} \left( u \right)$ and $\hat{\psi}(s)$
are Laplace transforms of the waiting time PDF.
Note that Eqs. (\ref{Eq17},\ref{Eq18}) give the proper normalization
since
$\sum_{n_a=0}^\infty \sum_{n_r=0} ^\infty P_{u,s} ( n_a , n_r) = 1/ ( u s)$.
Using Eq. (\ref{Eq18}) and 
$\langle n_a n_r \rangle_{u,s}=\sum_{n_a=0}^\infty \sum_{n_r = 0}^\infty n_a n_r P_{u,s}(n_a, n_r)$ we find 
\begin{equation}
 \langle n_a n_r \rangle_{u,s}=  { \left[ \hat{\psi}\left( s \right) - \hat{\psi}\left(u \right)\right] \hat{\psi}\left( u \right) \over
s \left( u - s \right) \left[ 1 - \hat{\psi}(u)\right]^2 \left[ 1 - \hat{\psi}(s) \right] }, 
\label{eqccc}
\end{equation}
and the  averages \cite{Margolin}
$\langle n_a \rangle _{u,s} =   \hat{\psi}(u) /\{u s [ 1 - \hat{\psi}(u)]\},$
\begin{equation}
\langle n_r \rangle _{u,s} = { \hat{\psi}(s) - \hat{\psi}(u) \over s \left( u - s \right) \left[ 1 - \hat{\psi}(u) \right]\left[ 1 - \hat{\psi}(s) \right]}.
\label{eqddd}
\end{equation}
In principle once the double Laplace inversion of Eqs. 
(\ref{eqccc},\ref{eqddd}) is made, we can
calculate the ${\rm FR}$ parameter.

 If the dynamics is Markovian, namely the 
waiting time PDF is exponential $\psi(t) = R \exp( - R t )$
\begin{equation}
\langle n_a n_r \rangle =\langle n_a \rangle \langle n_r \rangle =  R t_a R t_r,
\end{equation}
and   ${\rm FR}(t_a,t_r)=0$.
For any non-Markovian process with a non-exponential waiting time PDF
the ${\rm FR}$  parameter is generally not equal zero.  


If the average waiting time
 $\langle \tau \rangle =\int_0 ^\infty \tau  \psi(\tau) {\rm d} \tau $
is finite 
and in the limit $t_a \to \infty$ 
the fluctuation-response  parameter Eq. (\ref{Eq04}) 
satisfies
\begin{equation}
\lim_{t_a \to \infty} {\rm FR}(t_a,t_r) = 0,
\label{eqFRtai}
\end{equation}
and the correlations are lost in this limit. 
To see this use  Eq. (\ref{eqccc}) in the $u \to 0$ limit, the
small $u$ expansion $\hat{\psi}(u) \sim 1 - u \langle \tau \rangle$ to find
$\langle X_a ^2 X_a \rangle \sim h  a^3
{ 1 \over u^2 \langle \tau \rangle} {1 \over s^2 \langle \tau \rangle}$.
Interestingly this  result is
valid for any $s>0$, namely both for short and long
response times $t_r$. Hence when $t_a \to  \infty$ we find 
$\langle X_a ^2 X_r \rangle \sim h a^3
{ t_a \over \langle  \tau \rangle} 
{t_r \over \langle  \tau \rangle}$ 
and similar calculations for $\langle (X_a)^2 \rangle$,
and $\langle X_r \rangle$ complete the proof of 
Eq. (\ref{eqFRtai}).
If the aging period is much larger than the {\em finite}
time between jumps the response is not correlated with the
fluctuations, since the particles had enough time
to equilibrate. 

 A very  different behavior is found in  the common situation
\cite{BouchaudREV,Metzler,Weitz}
where the average waiting
time is infinite namely when $0< \alpha<1$ in Eq. 
(\ref{eqpower}).
Then using 
Eq. (\ref{eqccc}),
in the limit of small 
$s$ and $u$ 
\begin{equation}
\langle \left(X_{a}\right)^2 X_r \rangle \sim 
 h a^3 
{ 1 \over A^2 } { u^\alpha - s^\alpha \over (u - s ) u^{2 \alpha} s^{1 + \alpha}}, 
\label{eqFR1}
\end{equation}
where the expansion $\hat{\psi}(u) \sim 1 - A u^\alpha$ was used. 
Skipping the technical details, 
we analytically invert the double Laplace transform
in Eq. (\ref{eqFR1})
to the double time domain and find  
\begin{equation}
\langle \left( X_{a} \right)^2 X_r \rangle \sim 
 h a^3  {t_{r} ^{2 \alpha} \over A^2} g\left( {t_a \over t_r} \right),
\label{eqmain1}
\end{equation}
which is valid in the limit of long times $t_a$ and $t_r$.
The scaling function in Eq. (\ref{eqmain1})
is a hypergeometric function
\begin{equation}
g(x) = 
 {x^\alpha\  _2 F_1 \left[ 1 , - \alpha; 1 + \alpha; -x\right] \over \Gamma^2 \left( 1 + \alpha \right)}  - {x^{2 \alpha} \over \Gamma\left( 1 + 2 \alpha \right)}. 
\label{eqmain2}
\end{equation}
Eq. (\ref{eqmain1}) is a main result of this paper, since it shows
that even in the long aging time limit a non-trivial correlation between
fluctuation and response exists. 
The hypergeometric function in Eq. (\ref{eqmain2}) is tabulated in
Mathematica hence the solution is not a formal expression. 
Using
$\langle\left( X_a \right)^2 \rangle \sim a^2 { t_{a} ^\alpha \over A \Gamma( 1 + \alpha) }$, 
and the aging response of the model \cite{ACTRW}
$\langle X_r \rangle \sim h a { \left(t_{a} + t_r \right)^\alpha - t_{a} ^\alpha \over A \Gamma( 1 + \alpha) }$, 
the dimensionless fluctuation-response
parameter is
\begin{equation}
{\rm FR}(x) = 
{ _2 F_1\left( 1, - \alpha ; 1 + \alpha, - x \right)  -  x^\alpha {\Gamma^2\left( 1 + \alpha\right) \over \Gamma\left( 1 + 2 \alpha \right)}
 \over 
\left( 1 + x\right)^\alpha - x^\alpha } - 1,
\label{FRalpha}
\end{equation}
where $x=t_a /t_r$. If $\alpha=1$ we have ${\rm FR}(x) = 0$ indicating that 
the non-trivial correlations 
are found in the limit of long times,
only for anomalous processes with $\alpha < 1$. 
Eq. (\ref{FRalpha}) is valid in the limit $t_a \to \infty$ and 
$t_r \to \infty$ 
their ratio $x$ remaining finite. 

\begin{figure}
\begin{center}
\epsfxsize=80mm
\epsfbox{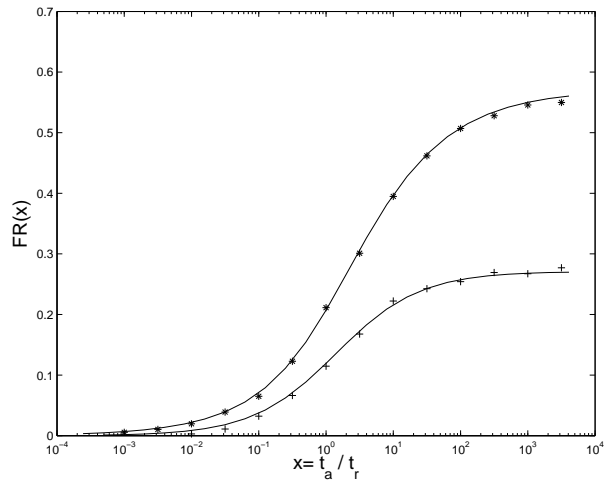}
\end{center}
\caption{   
The fluctuation-response parameter ${\rm FR}(x)$
 as a function of $x=t_a/t_r$.
Simulations and theory agree well without  fitting.
We use $\alpha=3/4$ (crosses) and $\alpha=1/2$ (stars) and see
that correlations are stronger when $\alpha =1/2$. 
}
\label{fig2}
\end{figure}

Comparison between simulations
of the CTRW process
and Eq. (\ref{FRalpha}) for $\alpha=1/2$ and $\alpha=3/4$ is made in Fig.
\ref{fig2}.
The figure illustrates that  the correlations between
fluctuations and response  becomes larger 
as $x= t_a/t_r$ is increased. 
This is the expected behavior, the larger is the aging time $t_a$
compared with the response time $t_r$ the stronger is the correlation,
since if $t_r>>t_a$ the particle already ``forgot'' its behavior
in the aging period. 
The Fig. also demonstrates that  
as $\alpha$ is decreased the correlations get stronger.
In the simulation
$t_a=3*10^6$, $\psi(\tau) = \alpha \tau^{-(1 + \alpha)}$ for $\tau>1$
otherwise it is zero.

Using Eq. 
(\ref{FRalpha})
we find in the limit
$x=t_a/t_r<<1$ 
\begin{equation}
{\rm FR}(x) \sim \left( 1 - { \Gamma^2 (1 + \alpha) \over \Gamma\left(1 + 2 \alpha\right)} \right) x^\alpha + O(x),
\label{eqsmallx}
\end{equation}
namely weak correlations
between fluctuations and response.
In the opposite limit, of the aging regime  of
$x>>1$ the correlations are stronger, and we find
\begin{equation}
{\rm FR}(x) \sim \left[ { \alpha | \Gamma(\alpha) |^2 \over \Gamma(2 \alpha)} - 1\right] - { 1 \over 1 + \alpha } { 1 \over x^\alpha} , \ \ x \to \infty.
\label{eqlargex}
\end{equation}
The leading term gives the non-trivial behavior of the fluctuation-response
parameter when $t_a/t_r \to \infty$. 
We find the bounds
$0 \le \lim_{x \to \infty} {\rm FR}(x) 
\le 1$
where the lower bound with zero correlations corresponds to $\alpha \to 1$ and 
the upper bound of strong correlations is found when $\alpha \to 0$.

  Applying linear response theory to Eq. (\ref{eqmain1}), 
yields the connection between the correlation function
and Physically observable parameters. 
In this same limit aging Einstein relations
between the ensemble average response and the fluctuations in the
absence of the field are valid \cite{ACTRW,Bertin}. We find
using $h=a F/ 2 k_b T \to 0$ 
\begin{equation}
\langle \left(X_a\right)^2 X_r \rangle \sim 
{ 2 F D_{\alpha} ^{2} t_{r} ^{2 \alpha} \over k_b T } g \left( { t_a \over t_r } \right),
\label{eqlast}
\end{equation}
where $D_\alpha=a^2/ 2 A$ is the fractional diffusion coefficient, 
which according to
its definition is 
$\langle X^2 \rangle \sim 2 D_{\alpha} t^\alpha/ \Gamma(1 + \alpha)$ 
\cite{Metzler,BarkaiPRE}. 
Eq. (\ref{eqlast}) is important since it shows
that the transport coefficient $D_\alpha$ and the exponent $\alpha$,
{\em  describing the
fluctuations in the absence of the external driving field}, 
are the only system parameters needed for the determination of the correlation
between fluctuations and the response.

\begin{figure}
\begin{center}
\epsfxsize=80mm
\epsfbox{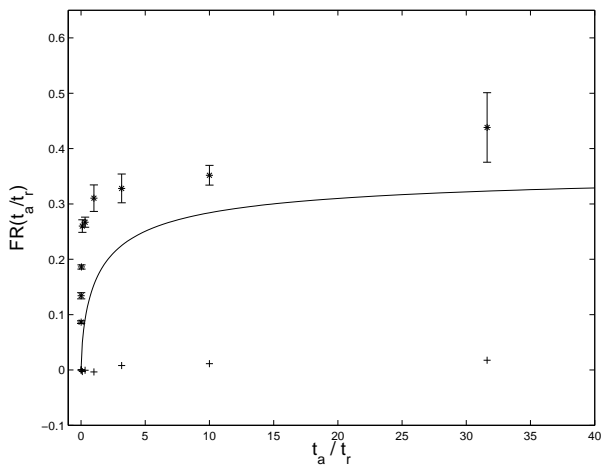}
\end{center}
\caption{ 
The ${\rm FR}$ parameter for the quenched trap model.
 The $+$ is for the high temperature
phase
$T= 5 T_g /2$ namely $T>T_g$ which does not exhibit correlations 
${\rm FR}\simeq 0$, while the stars are
for the aging phase $T=T_g/2$ which exhibits non-trivial
correlations.  The solid curve is the CTRW theory. 
}
\label{fig3}
\end{figure}

{\em Model 2} As shown by Feigelman and Vinokur \cite{First}
transport in disordered systems with quenched disorder
may exhibit aging effects. Hence it is natural to check if fluctuations
and response are correlated for models of quenched disorder, and
if so how do they compare with 
those we obtained analytically in the annealed CTRW model?
In particular do quenched models also exhibit a 
transition between an aging regime with strong correlations 
to a regime with vanishing correlations when $t_a \to \infty$?
For that aim we consider the quenched trap model 
on a one dimensional lattice \cite{Rinn,Bertin}. 
Each lattice site
$i$ has a  fixed random energy $E_i>0$, which is the energy
barrier the particle has to cross in order to jump from
$i$ to $i+1$ or $i-1$. The energy barriers are all
independent identically distributed random variables
with a common PDF $\rho(E)= (T_g)^{-1} e^{ - E/T_g}$.
 The PDF of escape times 
from site $i$ is exponential with a mean escape
time $\tau_i=exp(E_i/T)$. Notice that according to this Arrhenius
law small fluctuations in the energy may lead to exponentially
large fluctuations in the waiting time in site $i$.
After waiting in the trap for a random time the particle
has a probability $1/2$ of jumping left or right if the system
is not biased. It has a probability $(1\pm h)/2$ of jumping
left or right when the bias is not zero.
In simulations one lets the system evolve without
bias in the aging period, $(0,t_a)$ and then a bias is switched on. 

It is well known that the model exhibits aging behaviors
when $T<T_g$ \cite{WEB}.
Bertin and Bouchaud 
showed that the aging exhibits linear and non-linear  types of response,
depending
on the magnitude of $h$ \cite{Bertin}. Here 
we consider only the linear response regime
of $h \to 0$.
The trap model is not an exactly  solvable model 
since the effect of the quenched disorder is to induce non-independent
waiting times in the random walk. 
 Hence
we investigate the trap model using numerical simulations.  

  Simulation result for the ${\rm FR}$ parameter 
are shown in Fig.  \ref{fig3}. We choose  a parameter
set which is known to exhibit aging behavior \cite{Bertin}
$t_a=10^6$, $h=0.008$, and  vary
$t_r$. We choose two values of temperature  $T$ the first is in
the high temperature ergodic phase $T/T_g = 5/2$, in this case we
see that the ${\rm FR}$ parameter is practically zero with some small
deviations due to the finite time of simulation.
For $T/T_g=1/2$, namely in the aging phase, we see
strong correlations between the fluctuations and the response
especially when $t_r<t_a$. 
We also plot the ${\rm FR}(x)$ parameter for the CTRW model
using the exponent $\alpha= 2 T/T_g/(1 + T/T_g)=2/3$
($\alpha$ is the exponent describing
the averaged  response function
\cite{Bertin}).  The Fig.
clearly demonstrates that the ${\rm FR}$ parameter in the
CTRW theory
is smaller than the corresponding  ${\rm FR}$ parameter of the
quenched trap model.
In the quenched trap model,
 unlike  the  CTRW
process, the diffusion  is strongly
correlated in the sense that a particle once returning
to a specific trap will recall its waiting time
for that trap, hence the ${\rm FR}$ parameter for the
quenched model is larger than the one found for the CTRW.

 To conclude we showed that for any non-Markovian CTRW, 
correlations between the fluctuations in the
aging period  and the response are finite
if the aging period is finite. 
Both for the CTRW with $\alpha<1$ and for the quenched trap model
with $T<T_g$ a non-trivial ${\rm FR}$ parameter was found,
even in the limit of large aging times.  
These correlations are found in the phase where the models
exhibit aging transport, and hence we suspect that generally
systems which exhibit  aging may exhibit similar correlations
between fluctuations and response.  
The advance of single molecule tracking makes this research timely,
since we showed that the
response of individual particles to an external field depends
on their history of fluctuations.  

{\bf Acknowledgment} This work was supported by the 
Israel Science Foundation. I thank  J. P. Bouchaud
and G.  Margolin for their insight and discussions.

\end{document}